# Toward opportunistic coronary artery calcium assessment from dual-layer dual-energy chest radiography using deep learning: a simulation study

Linxi Shi[a*], Akyl Swaby[b], Kaitlin Hellier[b], Praveen Bharathi[b], Scott Hsieh[c], Dominik Fleischmann[a], Shiva Abbaszadeh[b], Adam S. Wang[a]

[a] Department of Radiology, Stanford University, Stanford, CA 94305, USA
[b] Radiological Instrumentation Laboratory, Electrical and Computer Engineering Department, University of California, Santa Cruz
[c] Department of Radiology, Mayo Clinic, Rochester, Minnesota, USA
*Corresponding Author: linxit@stanford.edu

## Abstract

**Background**: Coronary artery calcium (CAC) is a strong imaging biomarker of atherosclerotic burden and future cardiovascular events. While CT-based CAC scoring is clinically established, broader deployment for screening is limited by cost, radiation, and workflow considerations. Dual-energy (DE) chest radiography is widely accessible and may enable opportunistic CAC assessment if coronary calcifications can be reliably separated from overlapping bony anatomy.

**Purpose:** To develop and evaluate a multi-task deep learning framework for opportunistic CAC localization and CT-referenced score prediction using dual-layer DE chest radiography, and to assess the effects of projection angle, image representation, and exposure level under matched-dose training and testing conditions.

**Methods:** This retrospective simulation study used 550 calcium-scoring CT exams with CT-derived CAC reference masks and Agatston scores. Dual-layer DE chest radiographs were simulated using a physics-based detector model at multiple projection angles and exposure settings. A modified nnU-Net jointly predicted pixel-wise CAC localization and a CT-referenced CAC score from the continuous CAC probability map. AP (0°), 45°, and lateral (90°) views were compared, followed by lateral-view comparisons of single-energy (SE), DE bone, and high-pass filtered DE bone images. Separate dose-matched models were trained and tested at 70, 50, and 24 mAs. Data were split at the patient level (90% development, 10% independent testing), with five-fold cross-validation used for fold-averaged inference. Performance was assessed using Dice and score-based classification at CT CAC thresholds of >0, >100, >400, and >1000; AUC uncertainty was estimated by stratified bootstrap and paired AUC comparisons used DeLong tests with Holm-Bonferroni correction.

**Results:** Lateral projections showed the highest CAC visibility and model performance, with Dice increasing from 0.23 (AP) and 0.22 (45°) to 0.34 (lateral); the corresponding AUCs for CAC >100 were 0.81, 0.77, and 0.86. These AUC differences across projection angles were not statistically significant after multiple-comparison correction. At the lateral view, DE bone images outperformed SE radiographs (Dice: 0.33 vs. 0.19; AUC for CAC >100: 0.86 vs. 0.64), and the AUC advantage of DE bone over SE remained significant at CAC >100, >400, and >1000 after correction. High-pass filtering provided little additional score-classification benefit. Dice decreased with exposure, whereas no significant AUC differences were observed across the matched-dose models.

**Conclusions:** This simulation-based feasibility study supports lateral dual-layer DE bone imaging as a promising configuration for opportunistic CAC assessment. DE material decomposition provided a statistically supported advantage over SE radiography, while the lateral-view and dose analyses showed consistent performance trends that warrant validation in larger physical and clinical studies.

## 1.Introduction

Coronary artery calcium (CAC) is a well-established imaging biomarker of atherosclerotic burden and future cardiovascular events, and CAC scoring provides incremental risk stratification beyond traditional clinical predictors. In routine practice, CAC is most commonly quantified on non-contrast cardiac CT using the Agatston method, and contemporary prevention guidelines endorse CAC scores for selected patients when risk-based treatment decisions (e.g., statin initiation) remain uncertain[1–3].

Despite its clinical value, CT-based CAC screening is not universally deployed. Practical limitations include cost, radiation exposure, and availability. Opportunistic screening, such as extracting actionable cardiovascular risk information from imaging already obtained for other indications, offers an attractive pathway to broaden access without requiring a dedicated CT CAC exam. In this context, chest radiography (CXR) is particularly compelling due to its global ubiquity and throughput, but CAC detection on single-energy projection images is fundamentally challenging because small coronary calcifications are obscured by overlapping anatomy.

Dual-energy (DE) radiography aims to mitigate anatomic overlap by leveraging energy-dependent attenuation to generate material-separated images (e.g., bone and soft-tissue representations)[4]. Several studies have explored DE CXR for CAC assessment across phantom, simulation, and clinical paradigms. For example, experimental phantom work demonstrated feasibility of calcium detectability at low doses, while simulation and clinical evaluations highlighted the role of confounders such as motion artifacts and overlapping with other bony structures in real-world interpretation[5–7].

Particularly relevant studies by Hsieh et al[8] and Boswell et al[9] estimated diagnostic accuracy of DE CXR for CAC detection and demonstrated that the lateral orientation substantially improves performance relative to anterior-posterior (AP) imaging. Importantly for opportunistic deployment, they also emphasized practical limitations of DE acquisition in vivo: patient motion can induce misregistration between low- and high-energy images, and they noted that approaches such as faster switching, gating, or spectral detectors could reduce or eliminate misregistration artifacts[9–11].

Motivated by these considerations, dual-layer (DL) detectors offer a single-exposure DE solution by capturing low- and high-energy information simultaneously, improving robustness to motion compared with kV-switching approaches. This capability has been characterized in detail for a prototype DL flat-panel detector, with prior work noting that single-exposure DL DE imaging is inherently less sensitive to patient and system motion than kV-switching methods[12,13].

In parallel, deep learning has been demonstrated as a promising approach for detection and segmentation in complex imaging tasks. Self-configuring segmentation frameworks such as nnU-Net have lowered barriers to building strong task-adapted baselines, which is attractive for CAC localization where signal is subtle and confounded by structured background anatomy and noise[14,15].

In this work, we study DL DE-CXR as an opportunistic screening tool for CAC that can leverage radiographs acquired for routine care. By combining single-exposure DL DE acquisition with automated deep-learning segmentation, our goal is to improve CAC conspicuity and enable objective CAC detection/localization suitable for scalable opportunistic workflows, while systematically evaluating how acquisition geometry and dose influence performance.

## 2. Methods and Materials

### 2.1 Patient Cohort

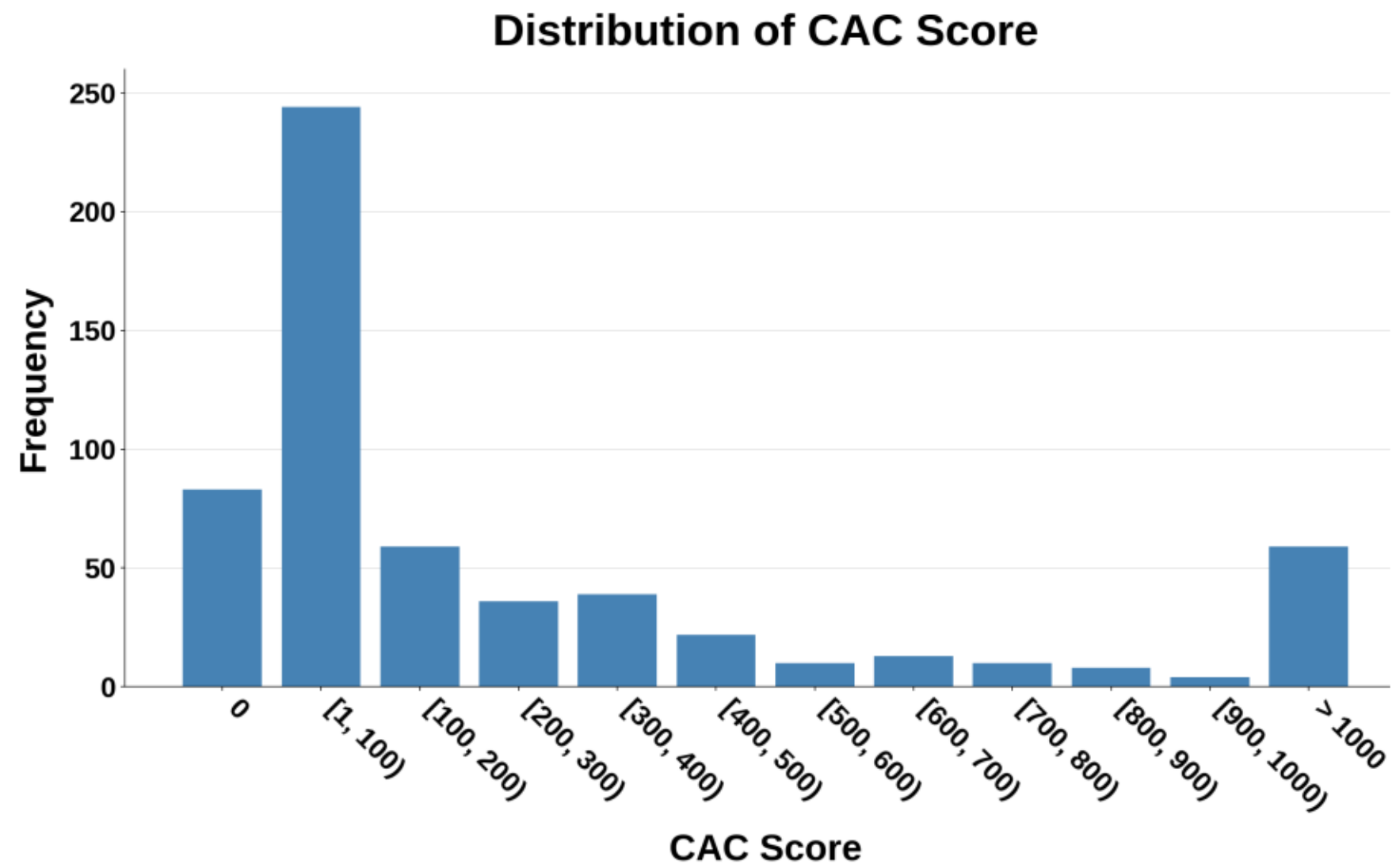

Figure 1. Histogram of CAC scores across patient cohort.

An IRB-approved retrospective collection of 550 calcium-scoring CT exams was used in this study, including exams from the UCLA Lundquist Institute and Stanford Health Care. Figure 1 shows the right-skewed distribution of CAC (Agatston) scores across the cohort. The median CAC score was 74 (IQR: 8–348), and 82 patients had CAC = 0. According to CAC score guidelines[16], low CAC scores (0 or <100) are associated with low cardiovascular risk, whereas higher CAC scores (>100) indicate increasing atherosclerotic burden and may prompt further diagnostic evaluation such as coronary CT angiography or functional testing. We prespecified CT CAC thresholds of >0, >100, >400, and >1000 for score-based classification, consistent with clinically relevant risk categories[17,18].

### 2.2 Simulated Dual-Energy CXR

The DE-CXR images were simulated from the CT dataset (Fig. 2). For each patient, the CAC reference was established from the clinical calcium-scoring CT reconstruction (3-mm slices, smaller FOV centered on the heart), while a large-FOV reconstruction (50 cm; 0.625–1.5 mm slice thickness) was used for DE-CXR simulation. The CAC masks were spatially mapped to the large-FOV CT grid. The CT-derived material-density volumes and CAC reference masks were then forward-projected using the same source-detector geometry, projection angle, and projection operator. No independent CT-to-radiograph registration was required; the nominal geometric registration error was therefore zero, with residual spatial uncertainty limited to resampling and finite voxel-to-detector discretization. Because projected CAC regions are small, even minor boundary shifts can disproportionately reduce Dice.

The first step was to conduct material decomposition on the full-size CT data, which was obtained using single-energy CT. The decomposition was performed based on the method by Wang et al[19] ,where the original CT number is converted into water and bone densities using thresholds of 1.2 g/cm$^3$ and 1.6 g/cm$^3$ as the linear transition region between pure water and pure bone  and calcium density scaling of 0.8. The resulting density maps, shown in Figure 2, represent the basis materials of water and calcium, which were then forward-projected onto a dual-layer flat-panel detector to generate low- and high-energy images. We assumed a geometry of DE CXR with a source-detector distance of 180 and a source-iso center of 160 cm. The dual-layer detector was simulated with a 43 × 43 cm$^2$ active area and pixel size of 0.6 × 0.6 mm$^2$. While the native pixel size of detectors is commonly around 0.15 mm, our capabilities to simulate small features are limited by the CT resolution. A polyenergetic spectrum from a 120 kVp source with 1 mm internal Al filtration was attenuated by the patient and then absorbed in a dual-layer (DL) detector. The DL detector was simulated with a top CsI scintillator thickness of 200 μm, a 1 mm Cu filter, and a bottom CsI scintillator thickness of 550 μm, matching a previously characterized prototype (Varex Imaging, Salt Lake City, UT).[12] The top and bottom projections, $I_{\text{top,init}}$ and $I_{\text{bot,init}}$, acquired by an energy integrating detector can be expressed as:

$$I_{\text{top,init}} = \int E \cdot I_{0,\text{top}}(E) e^{-t_w \mu_w(E) - t_b \mu_b(E)} dE, \tag{1}$$

$$I_{\text{bot,init}} = \int E \cdot I_{0,\text{bot}}(E) e^{-t_w \mu_w(E) - t_b \mu_b(E)} dE, \tag{2}$$

where I0,top(E) and I0,bot(E) are the effective spectra for the top and bottom layers as a function of E, tw and µw are the thickness and attenuation coefficient for basis material water, and tb and µb are those for basis material bone.

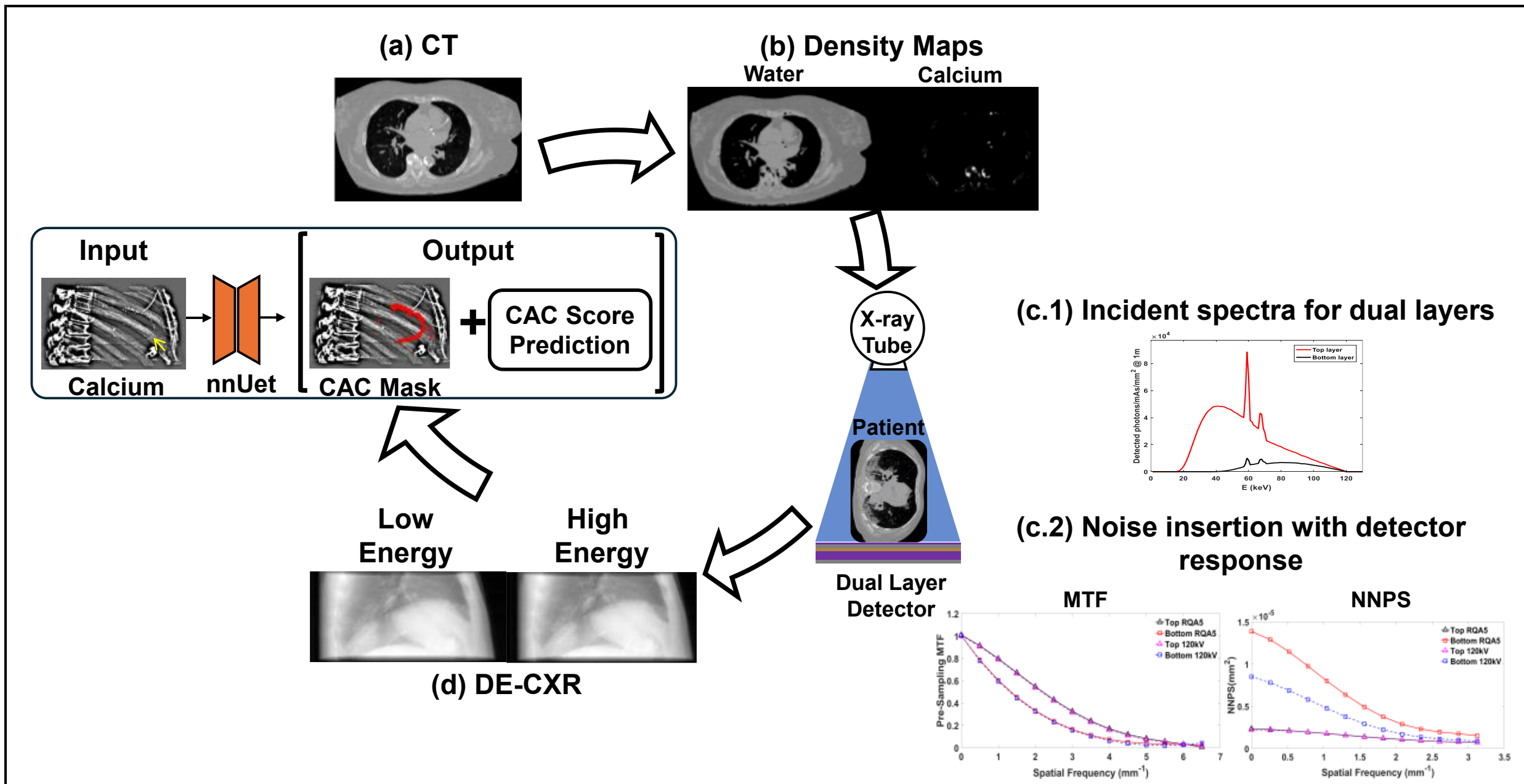

Figure 2. Workflow of using simulated DE-CXR for CAC detection.

The second step is to inject realistic noise into the projections. We need to know the detector modulated transfer function ($MTF$) and normalized noise power spectrum ($NNPS$) to accurately simulate the resolution and noise correlations. For this dual-layer prototype, we have previously characterized the detector response as published in Shi et al.[12], where the $MTF$ and $NNPS$ are described in the frequency domain. Let $K(\cdot)$ be the noise-shaping filter in the frequency domain, which relates to the measured $NNPS$ by $K(u,v) \propto \sqrt{NNPS(u,v)}$, where $u$, $v$ are the frequency coordinates along the $x$ and $y$ directions. The filter $K(u,v)$ is normalized to unity so that the magnitude of the injected variance is unchanged even after the noise correlations are introduced. Let $\mathcal{F}\{\cdot\}$ be the 2D Fourier transform and $\mathcal{F}^{-1}\{\cdot\}$ its inverse, we define:

$$\tilde{I}_{\text{top,init}}(u,v) = \mathcal{F}\{I_{\text{top,init}}(x,y)\}, \tag{3}$$

$$\tilde{I}_{\text{bot,init}}(u,v) = \mathcal{F}\{I_{\text{bot,init}}(x,y)\}, \tag{4}$$

$$\tilde{N}_{\text{top}}(u,v) = \mathcal{F}\left\{ \sigma\left(I_{\text{top,init}}(x,y)\right) \cdot w(x,y)\right\}, \tag{5}$$

$$\tilde{N}_{\text{bot}}(u,v) = \mathcal{F}\left\{ \sigma\left(I_{\text{bot,init}}(x,y)\right) \cdot w(x,y)\right\} \tag{6}$$

where the standard deviation map $\sigma(\cdot)$ is computed as:

$$\sigma(I_{\text{top,init}}) = \sqrt{\int E^2 \cdot I_{0,\text{top}}(E) e^{-t_w \mu_w(E) - t_b \mu_b(E)} dE}, \tag{7}$$

$$\sigma(I_{\text{bot,init}}) = \sqrt{\int E^2 \cdot I_{0,\text{bot}}(E) e^{-t_w \mu_w(E) - t_b \mu_b(E)} dE}, \tag{8}$$

Multiplying $\sigma(\cdot)$ with white noise $w(x,y) \sim \mathcal{N}(0,1)$ generates uncorrelated noise with the desired magnitude to inject, and the degree of correlation is introduced using filter $K(u,v)$. Thus, the noise injection process can be expressed as:

$$I_{\text{top}}(x,y) = \mathcal{F}^{-1}\{\tilde{I}_{\text{top,init}}(u,v)\, MTF_{\text{top}}(u,v) \; + \; \tilde{N}_{\text{top}}(u,v)\, K_{\text{top}}(u,v)\}, \tag{9}$$

$$I_{\text{bot}}(x,y) = \mathcal{F}^{-1}\{\tilde{I}_{\text{bot,init}}(u,v)\, MTF_{\text{bot}}(u,v) \;+\; \tilde{N}_{\text{bot}}(u,v)\, K_{\text{bot}}(u,v)\}. \tag{10}$$

$I_{top}$ and $I_{bot}$ are then log normalized to convert to line integrals $l_{top}$ and $l_{bot}$. Bone images were obtained by inputting the line integral pair, $l_{\text{top}}$ and $l_{\text{bot}}$, to a simulated material-dependent lookup table, a method applied in our previous study[20], which was simulated from pairs of water and calcium thicknesses ranging from 0 to 40 cm water and 0 to 100 mm calcium. The goal is to map line integrals from both layers to the corresponding paired material combination: $\{l_{\text{top}}, l_{\text{bot}}\} \rightarrow \{l_{\text{water}}, l_{\text{bone}}\}$. The detector model was experimentally constrained by the physical dual-layer prototype: detector geometry, scintillator thicknesses, inter-layer filtration, and measured MTF and NNPS were incorporated into the simulation. Scatter was not explicitly modeled; the present simulations therefore represent an idealized scatter-corrected condition.

### 2.3 Multi-Task Learning Model for CAC Segmentation and Scoring

#### 2.3.1 Model Infrastructure Modification

To enable simultaneous CAC localization and CAC score estimation from dual-energy chest radiographs, we implemented a multi-task learning (MTL) network that couples (i) a pixel-wise CAC segmentation branch with (ii) a case-level CAC score regression branch. The design follows a “shared backbone + task-specific heads” pattern to maximize feature sharing while allowing each task to specialize. Figure 3 shows the proposed MTL network infrastructure in implementation.

**i) Backbone segmentation network**

We used the nnU-Net as the core model to perform MTL. nnU-Net[14] is a UNet-based deep learning framework designed for medical image segmentation tasks, aiming to simplify model configuration and optimization while improving performance when processing highly diverse datasets. In this case, the network maps input DE CXR to CAC segmentation logits at full resolution. The network architecture follows a U-Net structure, featuring an encoder and a decoder interconnected by skip connections, and the feature extraction relies on plain convolutions. Downsampling is achieved using strided convolutions, while upsampling is accomplished through transposed convolution. Deep supervision is implemented by introducing auxiliary segmentation outputs at all decoding levels, except for the two lowest levels in the decoder. Given the 2D nature of CXR, the nnU-Net pipeline automatically selected 2D U-Net configurations, then adapts training properties according to the dataset fingerprint: In our configuration, the input patch size was 512 × 512 with batch size 12, using a base feature width of 32 channels and a maximum of 512 channels at the deepest level.

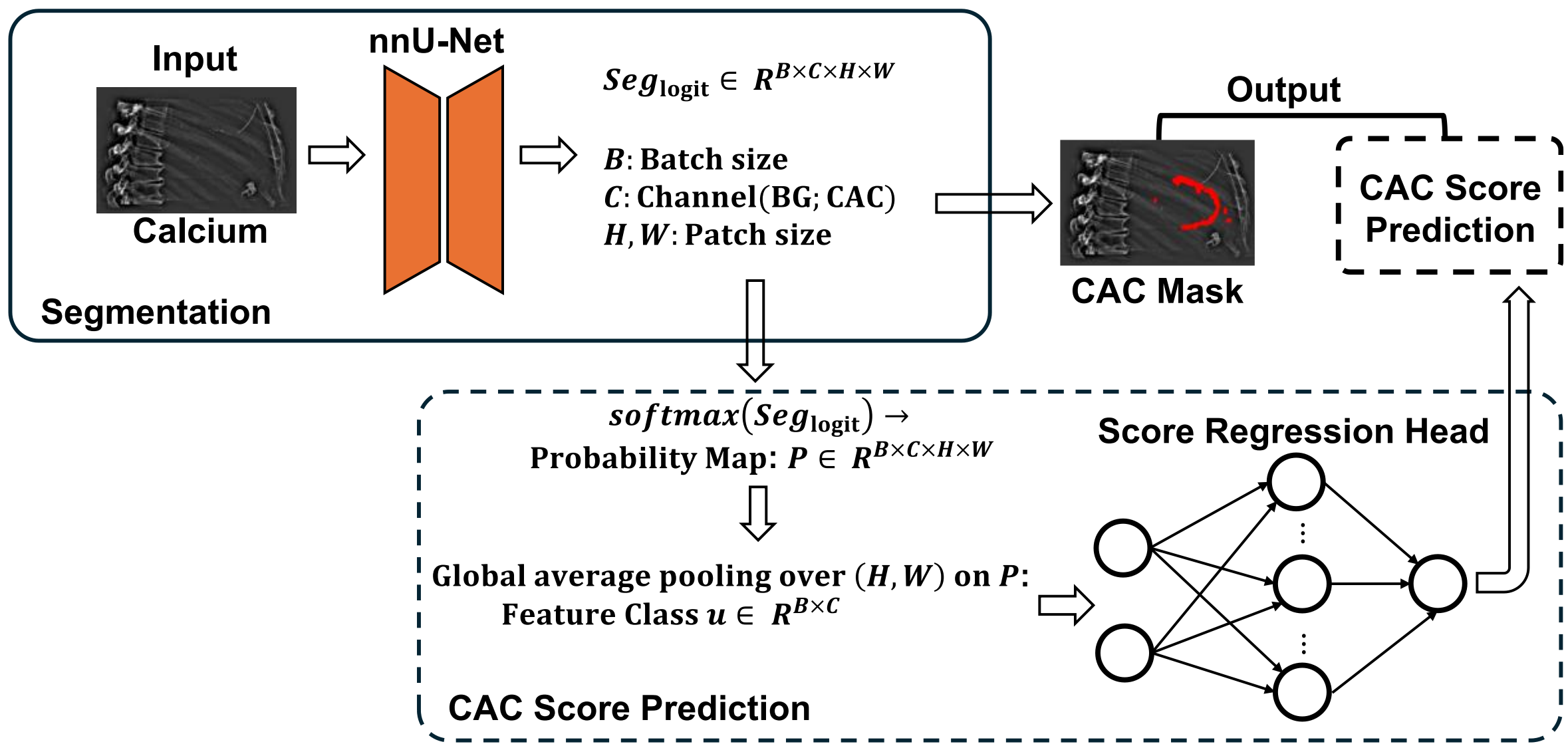

Figure 3. MTL network infrastructure for CAC segmentation and score prediction.

The network produces a dense segmentation output at full resolution. Let the segmentation logits be $S \in R^{B \times C \times H \times W}$, where $B$ is the batch size, $C$ corresponds to the two classes (background and CAC), and $H$ and $W$ corresponds to the training patch size. A softmax is applied along the class dimension to obtain probability maps:

$$P = softmax(S) \in R^{B \times C \times H \times W} \quad (11)$$

Let $P_{CAC}$ denote the continuous CAC probability map. A binary CAC mask was generated by thresholding $P_{CAC} > 0.5$ only for segmentation evaluation and visualization. The score-regression pathway used the continuous softmax probabilities directly, without thresholding.

**ii) Score regression head**

To obtain a case-level CT-referenced CAC score, the regression module uses the continuous softmax probability maps as input. Global average pooling is applied over the spatial dimensions to form a compact per-sample representation:

$$u \in R^{B \times 2}, \tag{12}$$

$$u_{B,C} = \frac{1}{H \cdot W} \sum_{i=1}^{H} \sum_{j=1}^{W} P_{B,C,i,j} \, , C \in \{\text{background}, \text{CAC}\} \tag{13}$$

Because the background and CAC softmax probabilities are complementary, the pooled two-class representation contains one independent quantity and can be interpreted as the spatially averaged CAC probability. This value is derived from the full nnU-Net response to the DE bone image and therefore represents aggregate learned CAC evidence rather than a thresholded fractional area. The pooled representation is passed through a two-layer MLP (2→32→1), which provides a nonlinear calibration from the bounded CAC-probability burden to the normalized CT-derived score. This formulation is not intended to explicitly reproduce the conventional Agatston area-and-density calculation; rather, it estimates a CT-referenced CAC score from the learned radiographic representation.

**iii) Joint multi-task loss function**

The CAC segmentation loss $L_{seg}$ follows standard nnU-Net practice using a combined loss of Dice loss ($L_{dice}$) and Cross-entropy loss ($L_{ce}$):

$$L_{seg} = L_{dice} + L_{ce} \tag{14}$$

The score-regression branch was trained on normalized CAC targets to improve optimization stability. Because the raw CAC distribution was strongly right-skewed and spanned a large dynamic range, z-score normalization was used to center and rescale the targets using the training-set mean and standard deviation.

$$y = \frac{s - \mu}{\sigma} \tag{15}$$

where $\mu$ and $\sigma$ denote the mean and standard deviation of the score distribution computed from the training set (and applied consistently to validation/test cases). During evaluation, predictions were mapped back to the original score domain via:

$$\hat{s} = \hat{y} \cdot \sigma + \mu \tag{16}$$

The score loss is the L1 loss between predicted and normalized target scores. L1 loss further limits the influence of extreme residuals because its gradient magnitude does not increase with error magnitude, unlike squared-error loss. Log-transformed targets were also explored but produced less stable optimization and poorer convergence in this dataset.

The overall training objective is a weighted sum of segmentation and score losses:

$$L = L_{seg} + \lambda \cdot L_{score} \tag{17}$$

where $\lambda$ controls the relative contribution of score regression to the total objective. In practice, $\lambda$ was chosen to be 0.02 to balance gradient scales so that neither task dominates optimization.

### 2.3.2 Training Scheme

To characterize the effect of acquisition geometry, we first simulated five views spanning AP (0°) to lateral (90°), including 22.5°, 45°, and 67.5°. All five views were used for the CAC-to-bone overlap analysis shown in Fig. 4. To avoid redundant training at closely spaced angles, three representative orientations, AP (0°), 45°, and lateral (90°), were selected for deep-learning evaluation. The CAC-to-bone fraction was used as an overall measure of anatomical separation, which was calculated as the summed projected CAC calcium density within the CT-derived CAC reference region divided by the summed total projected calcium/bone density within the same region. Higher values therefore indicate less contribution from overlapping bony anatomy.

After selecting the best-performing orientation, we evaluated an image-processing strategy intended to enhance CAC visibility. Following the approach proposed by Hsieh and Budoff[8], we applied high-pass filtering to the bone image by subtracting a blurred version of the original image, where the blur was generated using a Gaussian filter (σ = 5 pixels). We then assessed whether this enhanced CAC conspicuity translated into improved segmentation and score prediction performance. In parallel, we performed a comparison against single-energy (SE) CXR at the selected orientation to quantify the incremental benefit of DE material decomposition for CAC detection. Both datasets were simulated at the same dose setting (120 kVp, 70 mAs).

After selecting the lateral orientation, dose-dependent performance was evaluated at 70, 50, and 24 mAs[21]. A separate model was trained, validated, and tested at each exposure level; the patient split, network architecture, training procedure, and held-out test cohort were otherwise identical. Thus, this experiment evaluates matched-dose performance rather than cross-dose generalization of a single model. The simulated air kerma using Spektr at isocenter was approximately 0.05 mGy/mAs at the 160-cm source-to-isocenter distance, corresponding to approximately 3.5, 2.5, and 1.2 mGy at 70, 50, and 24 mAs, respectively. The longer source-to-isocenter distance relative to a typical CAC CT geometry (~61 cm) corresponds to an approximately 6.88-fold geometric reduction in incident fluence by the inverse-square law; this factor describes geometry only and is not a complete patient-dose comparison.

The dataset was randomly split at the patient level, with 90% used for model development and 10% held out for independent testing. The fixed test cohort contained 59 patients. For the prespecified thresholds, the positive/negative counts were 49/10 for CAC >0, 28/31 for CAC >100, 15/44 for CAC >400, and 5/54 for CAC >1000. The same held-out patients were used for all projection-angle, image-representation, and dose comparisons. CT-derived CAC reference masks were forward-projected to provide paired spatial supervision for the simulated radiographs.

Within the 90% development set, model selection was performed using 5-fold cross-validation. In each fold, 80% of the development set was used for training and 20% for validation, with patient cohorts re-sampled across folds to evaluate robustness across different subsets of the data. Each fold was trained for 100 epochs, and cross-validation results were used to summarize performance.

Evaluation on the 10% testing dataset used the average inference result across all the five folds for each testing scenario.

#### **2.3.3** Statistical Analysis

AUC uncertainty was estimated using 10,000 patient-level stratified bootstrap resamples of the held-out test cohort; the 2.5th and 97.5th percentiles were reported as 95% confidence intervals. Because the same patients were evaluated across imaging conditions, paired DeLong tests were used for predefined AUC comparisons. Holm-Bonferroni correction was applied within each experimental comparison family, with adjusted $p < 0.05$ considered statistically significant. Sensitivity and specificity were reported descriptively at the operating point maximizing the Youden index, while Dice was used as a descriptive measure of pixel-level localization.

## 3. Results

### 3.1 Optimal angle selection

Figure 4 provides a quantitative visibility rationale for view selection. As the projection angle increases from AP (0°) toward lateral (90°), the CAC-to-bone fraction increases, indicating progressively reduced overlap between coronary calcifications and high-attenuation bony anatomy. For the representative case in Fig. 4(a), the CAC fraction increases from approximately 0.10 (AP) to 0.40 (lateral) (Fig. 4(b)). At the cohort level, the average CAC fraction increases from approximately 0.18 (AP) to 0.36 (lateral) (Fig. 4(c)). This systematic increase is also visually apparent in the bone images in Fig. 4(a), where CAC is less embedded within rib/spine structures in the lateral projection.

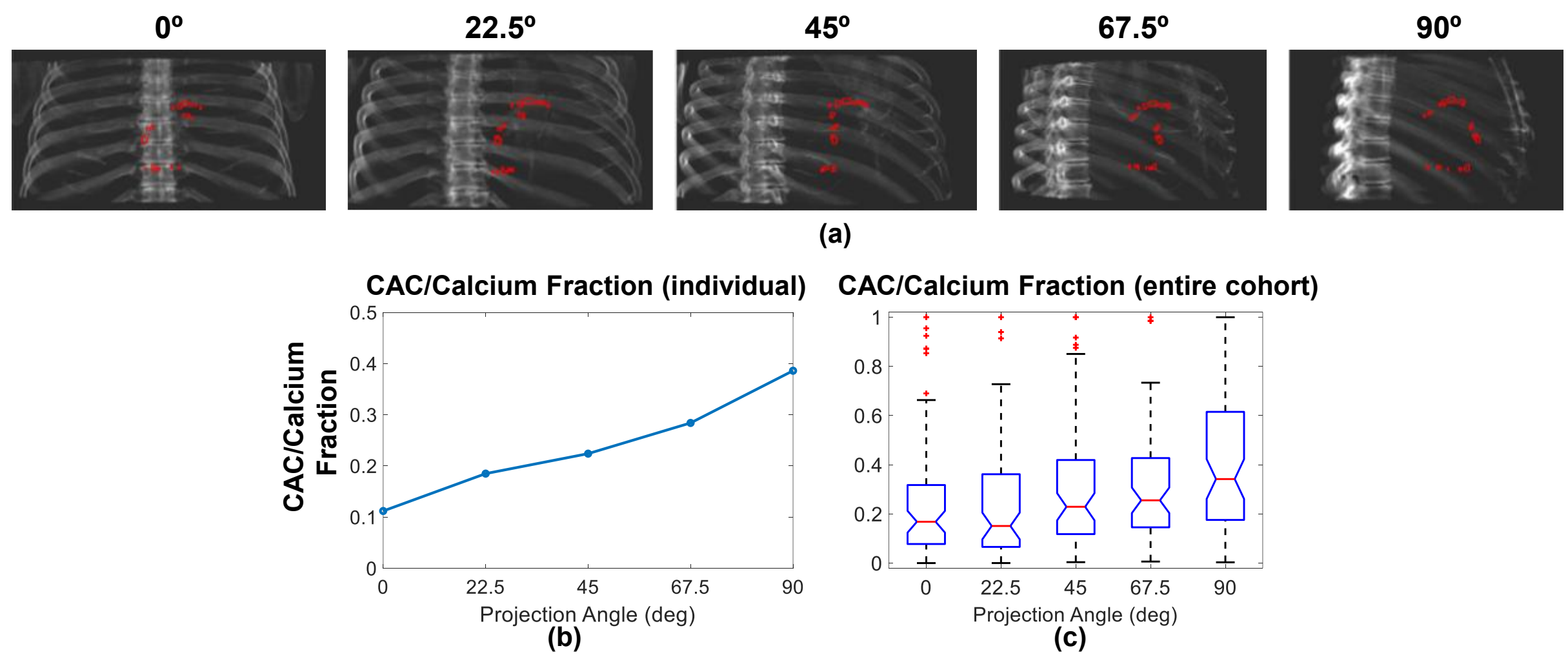

Figure 4. Effect of projection angle on CAC visibility. (a) Representative DE bone images across projection angles with the CT-derived CAC reference overlaid. The CAC/Calcium fraction is defined as the fraction of projected calcium signal within the CAC region relative to the total calcium/bone signal, with higher values indicating reduced overlap with surrounding bony anatomy. (b) CAC fraction as a function of projection angle for the representative case in (a), and (c) averaged across the cohort.

On the held-out test set, mean Dice scores were 0.23 (AP), 0.22 (45°), and 0.34 (lateral). AP and 45° predictions were more frequently fragmented or missed where CAC overlapped high-contrast bony edges, whereas the lateral view more often preserved contiguous CAC regions. Although the cohort-level CAC-to-bone fraction increased slightly from AP to 45° (20% to 23%), this modest change was not sufficient to improve Dice. Considerable overlap with ribs, vertebral edges, and the scapula remained at 45°, indicating that the fraction metric is an overall measure of separation and does not fully capture the spatial pattern of anatomical interference.

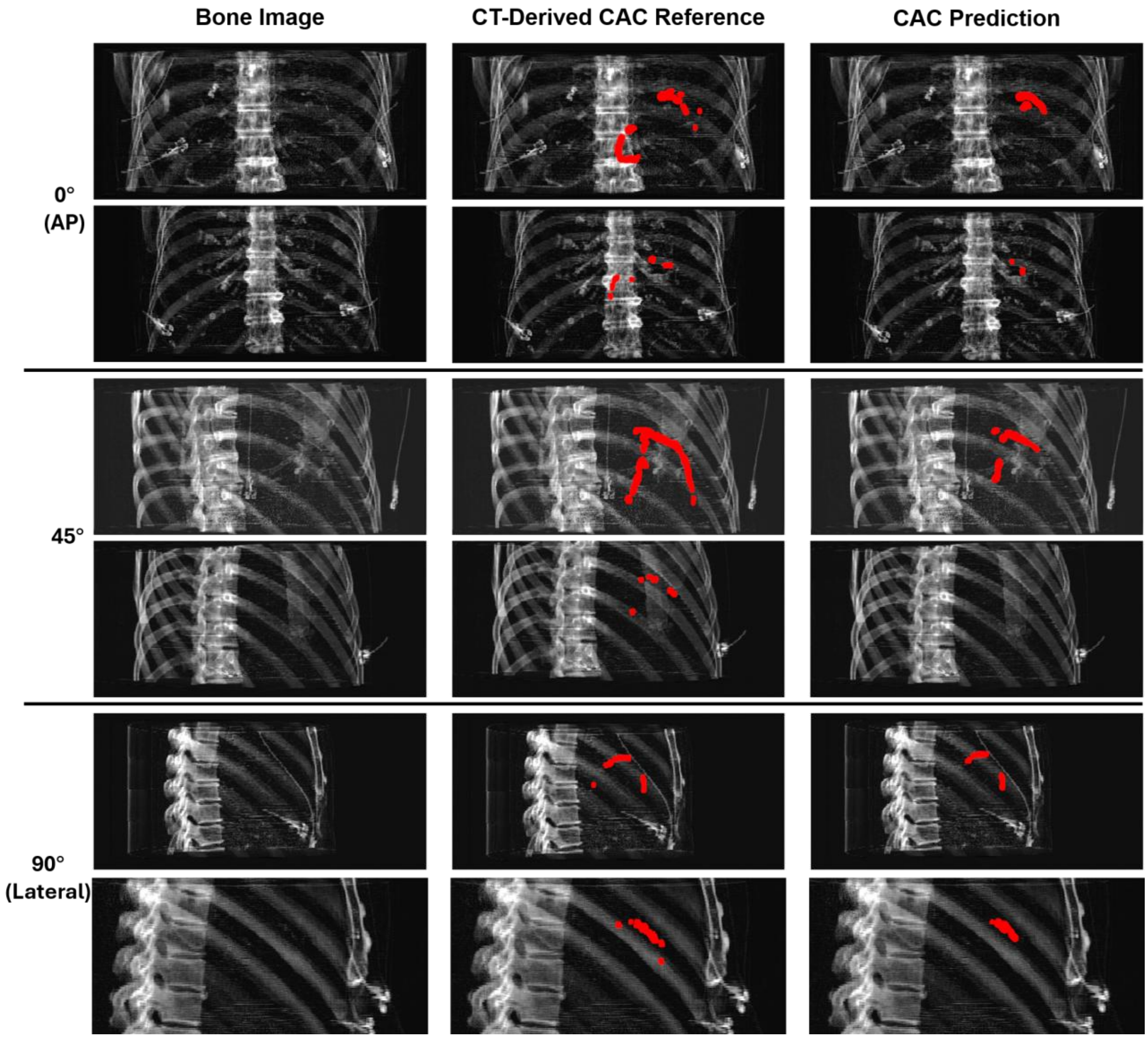

Figure 5. Qualitative comparison of CAC localization across projection angles. Columns show the simulated DE bone image, CT-derived projected CAC reference, and model-predicted CAC segmentation. Two representative test cases are shown for AP (0°), 45°, and lateral (90°) views.

Score-based discrimination showed a similar numerical trend (Fig. 6; Table 1). For CAC >100, AUCs were 0.81 (AP), 0.77 (45°), and 0.86 (lateral), with sensitivity/specificity of 0.86/0.65, 0.68/0.84, and 0.75/0.87, respectively. AUC was used as the primary inferential classification metric because it is independent of a single operating point; 95% bootstrap confidence intervals are reported in Table 1, where the best AUC performance for CAC >100 is in bold.

At CAC >400, AUCs were 0.87 (AP), 0.85 (45°), and 0.96 (lateral); at CAC >1000 they were 0.96, 0.79, and 0.98, respectively. Although the lateral projection consistently showed the highest or near-highest AUC, paired DeLong comparisons did not reach statistical significance after Holm-Bonferroni correction (adjusted $p > 0.05$). These differences are therefore interpreted as performance trends, supported by the overlap and localization analyses rather than as statistically significant AUC improvements.

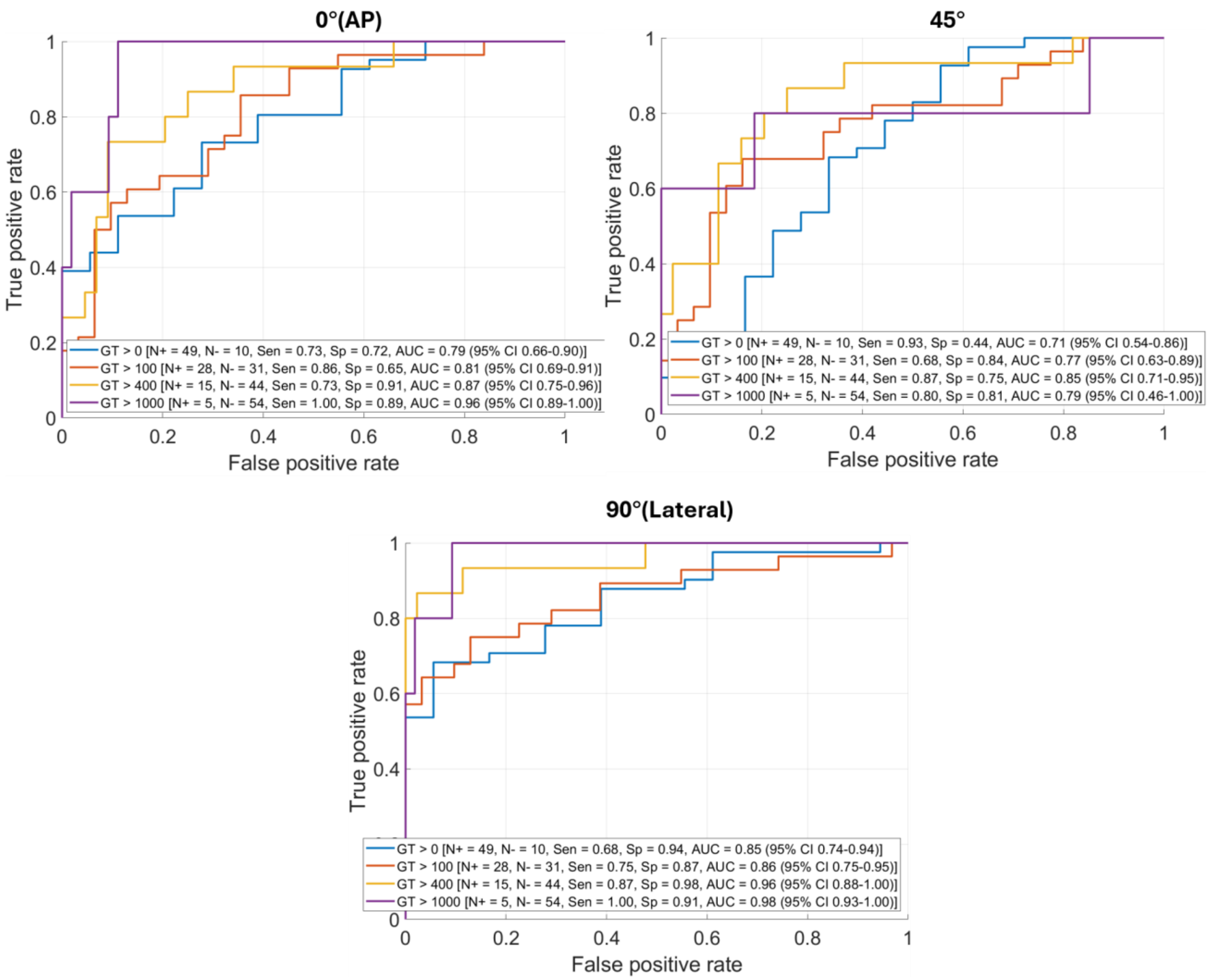

Figure 6. ROC analysis of CT-referenced CAC score classification across AP (0°), 45°, and lateral (90°) projections for CAC thresholds >0, >100, >400, and >1000.

**Table 1. Diagnostic accuracy for the four CAC thresholds across projection angles. AUC values are reported with 95% bootstrap confidence intervals.**

| Threshold | Sensitivity | | | Specificity | | | AUC of ROC (95% CI) | | |
|---|---|---|---|---|---|---|---|---|---|
| | 0° | 45° | 90° | 0° | 45° | 90° | 0° | 45° | 90° |
| >0 | 0.73 | 0.93 | 0.68 | 0.72 | 0.44 | 0.94 | 0.79(0.66-0.90) | 0.71(0.54-0.86) | 0.85(0.74-0.94) |
| >100 | 0.86 | 0.68 | 0.75 | 0.65 | 0.84 | 0.87 | 0.81(0.69-0.91) | 0.77(0.63-0.89) | **0.86(0.75-0.95)** |
| >400 | 0.73 | 0.87 | 0.87 | 0.91 | 0.75 | 0.98 | 0.87(0.75-0.96) | 0.85(0.71-0.96) | 0.96(0.88-1.00) |
| >1000 | 1.00 | 0.80 | 1.00 | 0.89 | 0.81 | 0.91 | 0.96(0.89-1.00) | 0.79(0.46-1.00) | 0.98(0.93-1.00) |

### 3.2 Calcium visibility enhancement

In this section, we evaluated image-processing strategies aimed at enhancing calcium conspicuity and examined whether improved visibility translated into better model performance. Following the observer-study workflow proposed by Hsieh and Budoff, we compared three input representations at the selected lateral (90°) view under a high-dose condition (70 mAs): single-energy (SE) CXR, DE bone image, and a high-pass filtered bone image (HP bone). Representative examples are shown in Fig. 7. Compared with SE, the DE bone image provides improved structural separation that increases the relative visibility of calcified regions. The high-pass filtered bone image further emphasizes high-frequency content, such as edges and small structures, but also visually accentuates noise and rib/spine texture. An example in Fig. 8 illustrates a case where CAC was correctly detected in both the bone and HP bone images with similar accuracy but was missed in the SE radiograph.

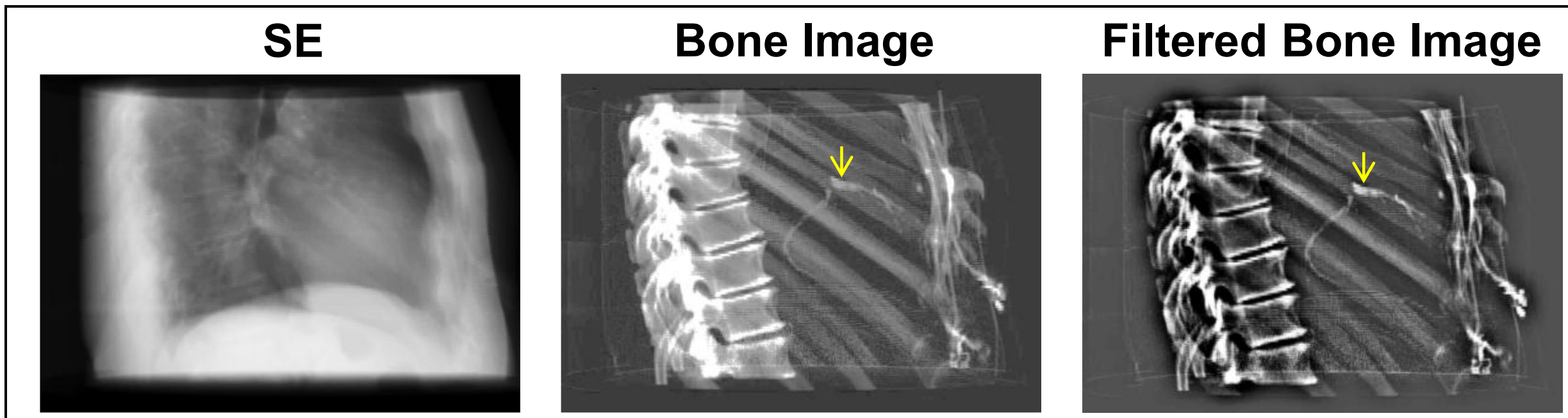

Figure 7. Lateral-view for calcium visibility enhancement experiments at high dose (70 mAs). Left to right: single-energy (SE) CXR, dual-energy (DE) bone image, and high-pass filtered bone image generated by subtracting a Gaussian-blurred version of the bone image. CAC is indicated by yellow arrow.

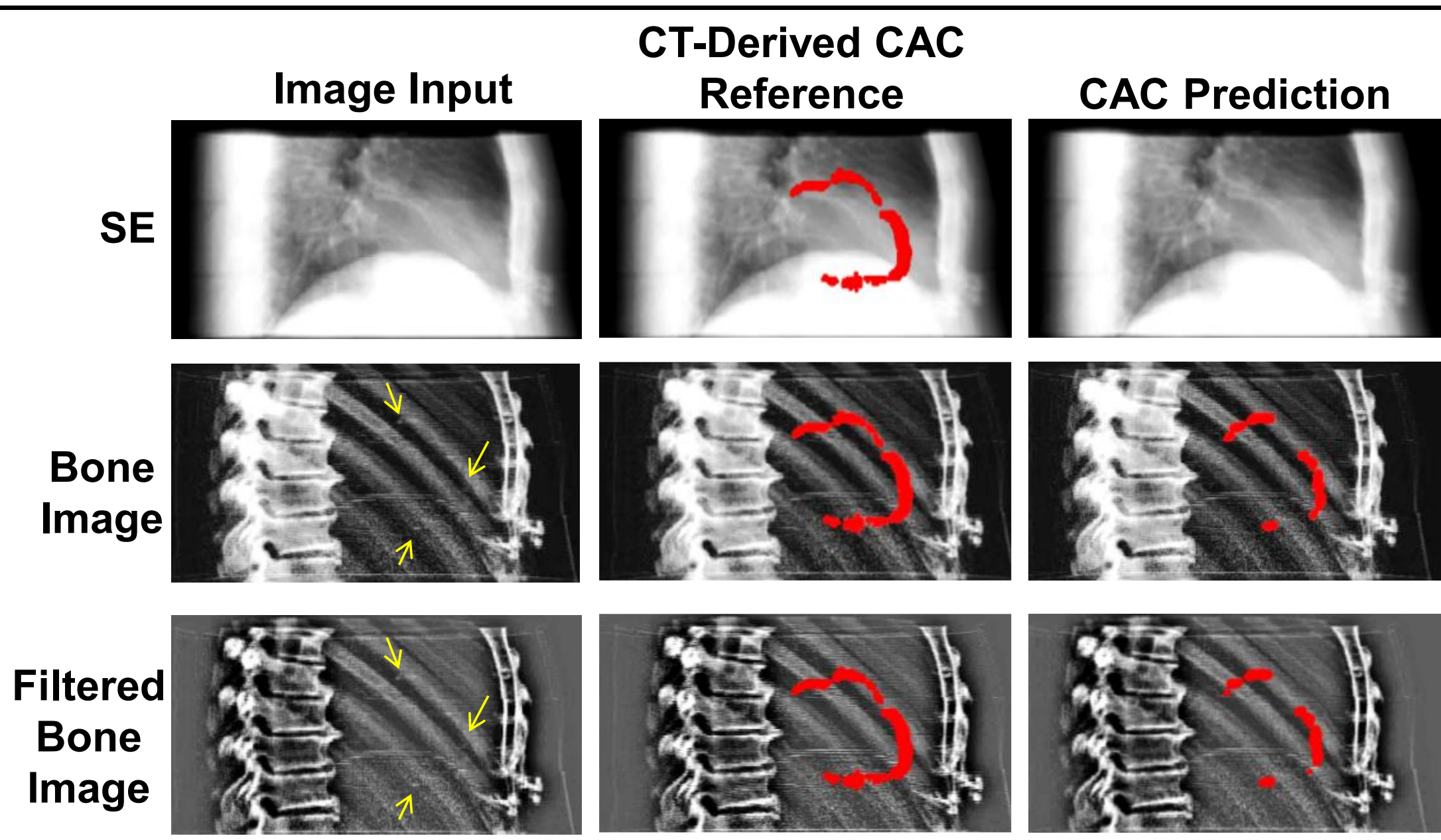

Figure 8. CAC segmentation results for the same case using SE, DE bone, and high-pass filtered DE bone images. For each row, the left column shows the image input, the middle column shows the CT-derived CAC reference, and the right column shows the model-predicted CAC segmentation. SE CXR often misses CAC due to soft-tissue overlap, while Bone and HP bone perform similarly at high dose.

Segmentation performance on the held-out test set was higher for the DE-derived representations than for SE radiographs. Mean Dice scores were 0.19 (SE), 0.33 (DE bone), and 0.34 (HP bone). The small difference between bone and HP bone indicates limited additional localization benefit from high-pass filtering at 70 mAs.

Score-based discrimination results are summarized in Fig. 9 and Table 2. DE bone images substantially outperformed SE radiographs at clinically relevant thresholds, with AUC increasing from 0.64 to 0.86 for CAC >100, from 0.58 to 0.96 for CAC >400, and from 0.58 to 0.98 for CAC >1000. Paired DeLong testing confirmed that the DE-bone versus SE differences at CAC >100, >400, and >1000 remained significant after Holm-Bonferroni correction. In contrast, DE bone and HP bone did not differ significantly at these thresholds (adjusted $p > 0.05$), indicating limited additional classification benefit from high-pass filtering.

For CAC >0, AUCs were 0.78 (SE), 0.85 (DE bone), and 0.87 (HP bone). Overall, the results support the benefit of DE material separation over SE radiography, while showing little incremental score-classification benefit from high-pass filtering.

### 3.3 Dose-dependent performance

To evaluate performance across exposure levels, separate dose-matched models were trained and tested at 70, 50, and 24 mAs using the same patient split. Bone and high-pass (HP) bone inputs were compared at each exposure. Lower exposure increased quantum noise and reduced the visibility of subtle calcifications, particularly in HP-filtered images where high-frequency noise and structured bony texture were emphasized (Fig. 10).

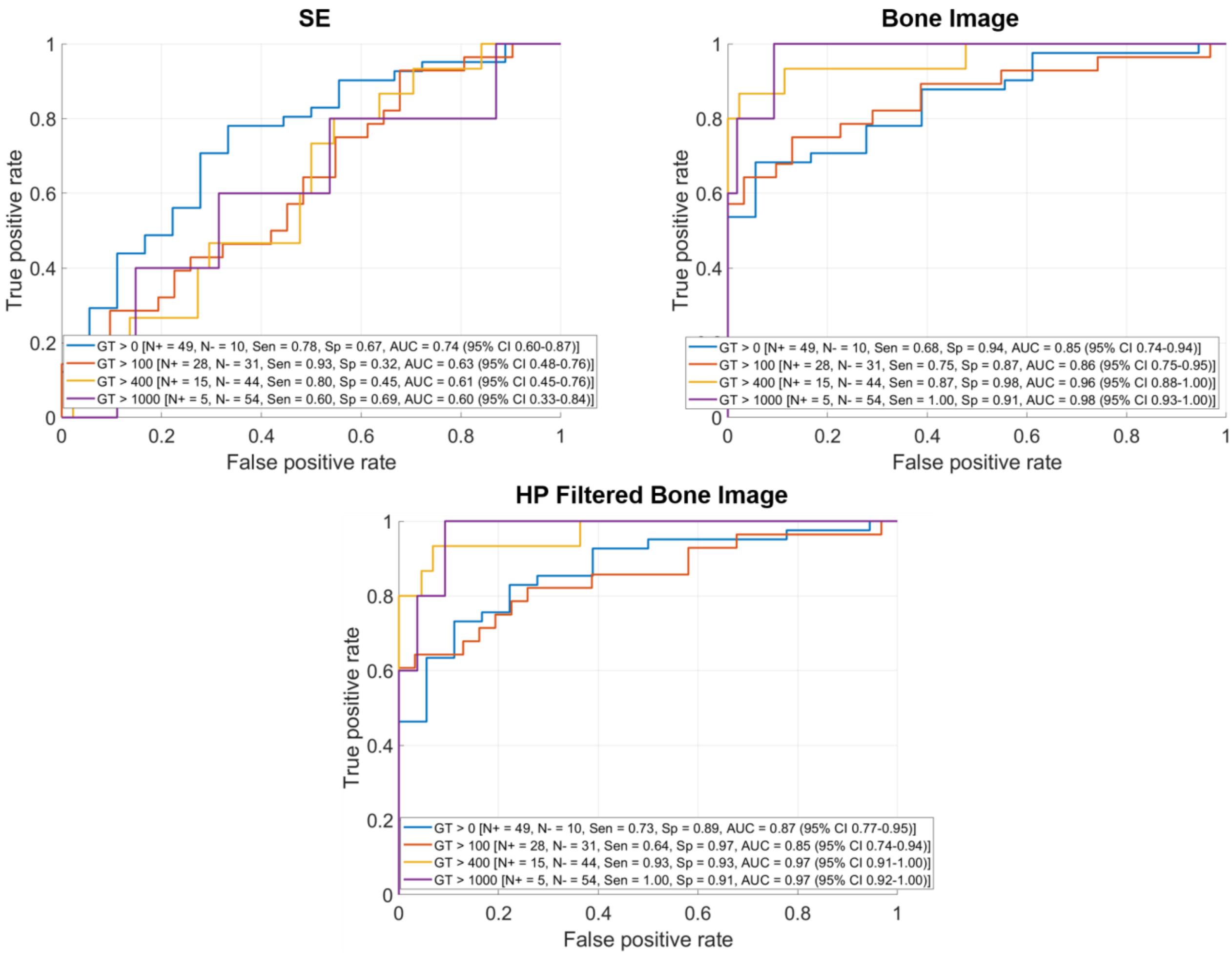

Figure 9. ROC analysis of CT-referenced CAC score classification for lateral-view SE, DE bone, and high-pass filtered DE bone inputs at CAC thresholds >0, >100, >400, and >1000.

**Table 2. Diagnostic accuracy for the four CAC thresholds across image representations. AUC values are reported with 95% bootstrap confidence intervals.**

| Threshold | Sensitivity | | | Specificity | | | AUC of ROC(95% CI) | | |
|---|---|---|---|---|---|---|---|---|---|
| | SE | Bone Image | Filtered Bone Image | SE | Bone Image | Filtered Bone Image | SE | Bone Image | Filtered Bone Image |
| >0 | 0.78 | 0.68 | 0.73 | 0.67 | 0.94 | 0.87 | 0.74(0.60-0.87) | 0.85(0.74-0.94) | 0.87(0.77–0.95) |
| >100 | 0.93 | 0.75 | 0.64 | 0.32 | 0.87 | 0.97 | 0.63(0.48-0.76) | **0.86(0.75-0.95)** | **0.85(0.74–0.94)** |
| >400 | 0.80 | 0.87 | 0.93 | 0.45 | 0.98 | 0.93 | 0.61(0.45-0.76) | 0.96(0.88-1.00) | 0.97(0.91–1.00) |
| >1000 | 0.60 | 1.00 | 1.00 | 0.69 | 0.91 | 0.91 | 0.60(0.33-0.84) | 0.98(0.93-1.00) | 0.97(0.92–1.00) |

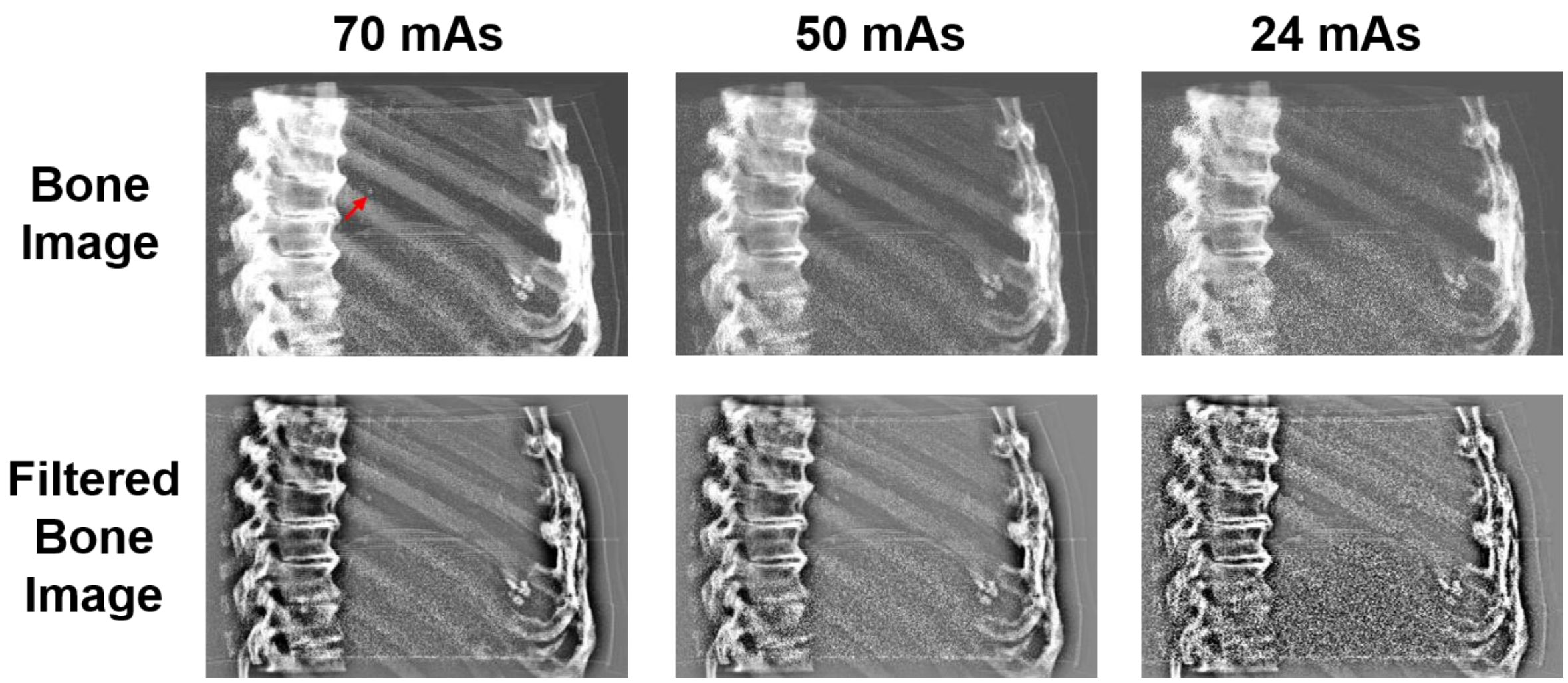

Figure 10. Representative lateral-view DE inputs at decreasing exposure levels. Top row: DE bone images; bottom row: high-pass filtered DE bone images. The red arrow indicates a small calcification.

### 3.3.1 Visual trend and segmentation results

Figure 10 illustrates that both bone and HP bone images degrade as the exposure level decreases; however, the high-pass filter further accentuates noise and fine rib texture at lower mAs. This observation supports the hypothesis that, although high-pass filtering can enhance calcification edges, it also raises the effective noise floor and amplifies structured anatomical clutter - factors that can impair pixel-wise segmentation and destabilize score estimation when the case-level score is derived from the predicted probability distribution.

Segmentation performance decreased with exposure for both representations (Fig. 11). Mean Dice for bone images decreased from 0.33 at 70 mAs to 0.30 at 50 mAs and 0.28 at 24 mAs; HP-bone Dice changed from 0.34 to 0.31 and 0.25. Because projected CAC occupies only a few detector pixels, small boundary errors can substantially reduce Dice; this effect becomes more pronounced as noise obscures CAC boundaries.

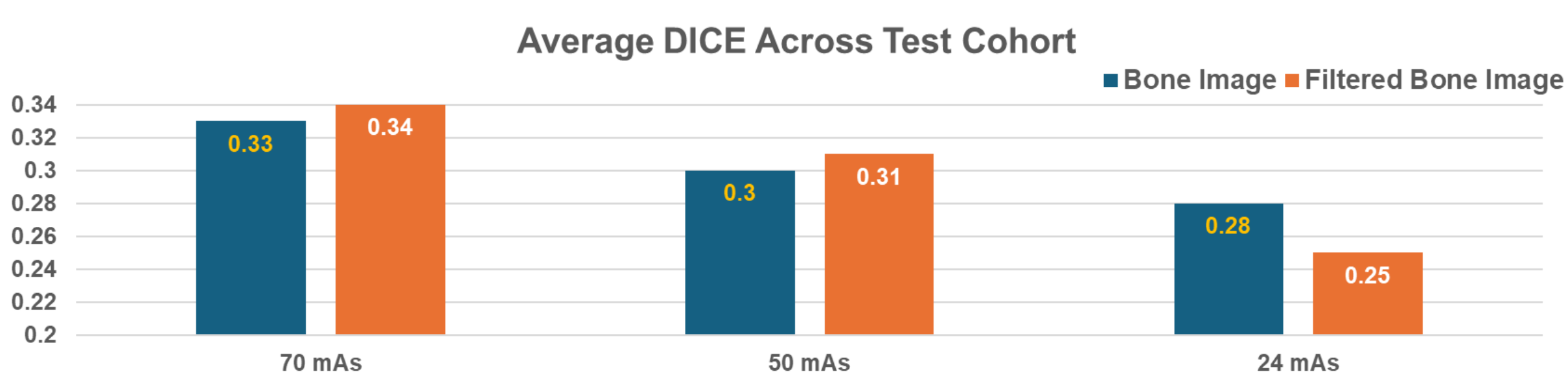

Figure 11. Mean Dice score across the held-out test cohort for DE bone and high-pass filtered DE bone inputs at 70, 50, and 24 mAs.

### 3.3.2 Score prediction performance across dose

ROC curves for the predicted CT-referenced CAC score are summarized in Fig. 12, with quantitative metrics and 95% bootstrap confidence intervals in Table 3. Because a separate dose-matched model was trained for each condition, these results characterize achievable performance at each exposure rather than cross-dose robustness of a single model.

Score-based discrimination changed only modestly across exposure levels. For CAC >100, bone-image AUCs were 0.86, 0.83, and 0.81 at 70, 50, and 24 mAs, respectively; HP-bone AUCs were 0.85, 0.84, and 0.82. Paired DeLong comparisons showed no significant AUC differences between 70 and 50 mAs or between 70 and 24 mAs at CAC >100, >400, or >1000 after multiple-comparison correction (adjusted $p > 0.05$). Thus, under matched-dose training and testing, case-level classification remained relatively stable even as pixel-level Dice declined.

This divergence reflects the different tasks: segmentation requires accurate pixel-level delineation, whereas case-level score prediction aggregates evidence across the image and is less sensitive to small local boundary errors.

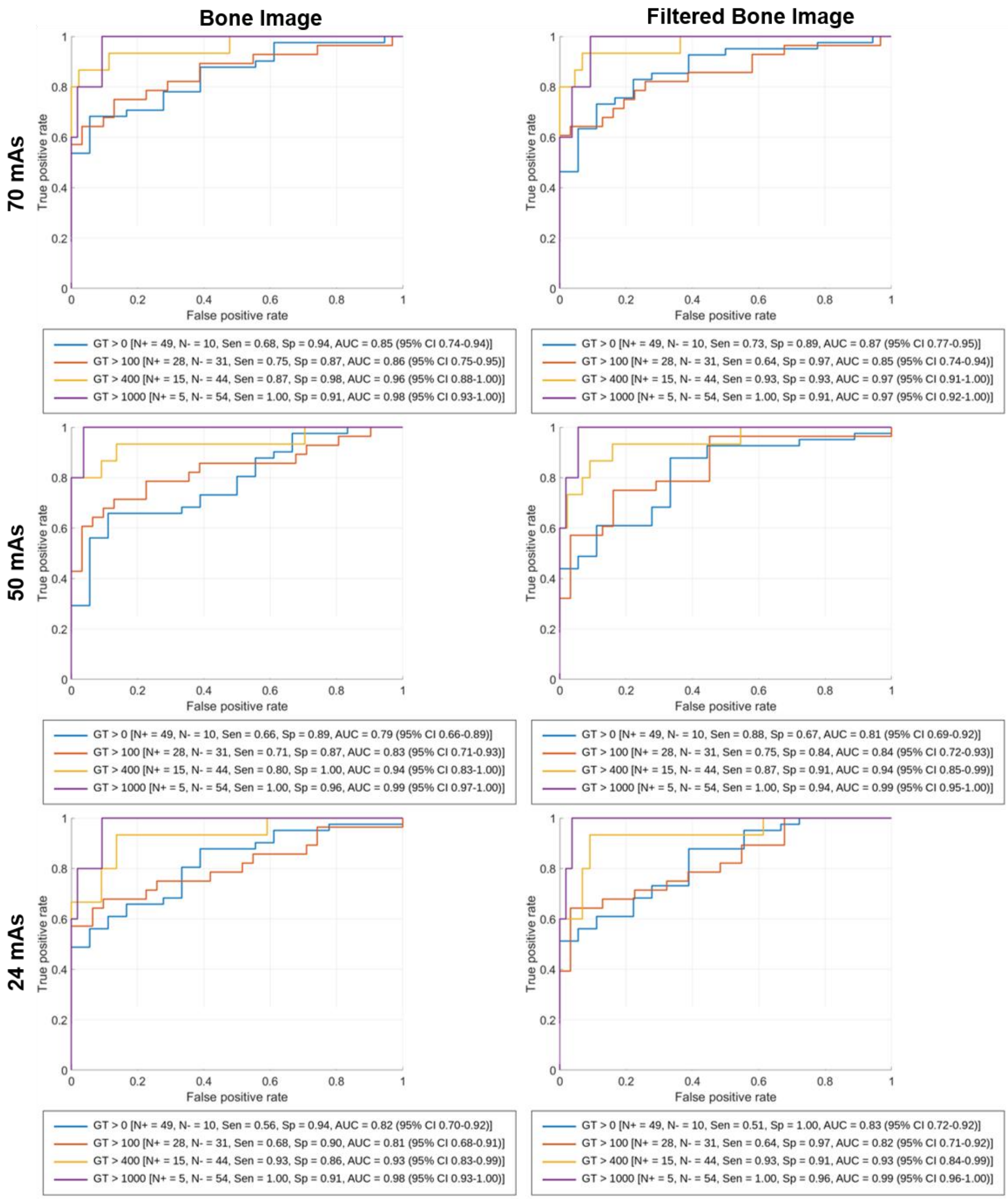

Figure 12. ROC curves for CT-referenced CAC score classification across exposure levels (70, 50, and 24 mAs) for DE bone images (left column) and high-pass filtered DE bone images (right column) at the lateral view.

**Table 3. Diagnostic accuracy for the four CAC thresholds across exposure levels. AUC values are reported with 95% bootstrap confidence intervals.**

| | Threshold | Sensitivity | | | Specificity | | | AUC of ROC(95% CI) | | |
|---|---|---|---|---|---|---|---|---|---|---|
| | | 70 mAs | 50 mAs | 24 mAs | 70 mAs | 50 mAs | 24 mAs | 70 mAs | 50 mAs | 24 mAs |
| Bone Images | >0 | 0.68 | 0.66 | 0.56 | 0.94 | 0.89 | 0.94 | 0.85 (0.74–0.94) | 0.79 (0.66–0.89) | 0.82 (0.70–0.92) |
| | >100 | 0.75 | 0.71 | 0.68 | 0.87 | 0.87 | 0.90 | **0.86 (0.75–0.95)** | **0.83 (0.71–0.93)** | **0.81 (0.68–0.91)** |
| | >400 | 0.87 | 0.80 | 0.93 | 0.98 | 1.00 | 0.86 | 0.96 (0.86–1.00) | 0.94 (0.83–1.00) | 0.93 (0.83–0.99) |
| | >1000 | 1.00 | 1.00 | 1.00 | 0.91 | 0.96 | 0.91 | 0.98 (0.93–1.00) | 0.99 (0.97–1.00) | 0.98 (0.93–1.00) |
| Filtered Bone Images | >0 | 0.73 | 0.88 | 0.51 | 0.89 | 0.67 | 1.00 | 0.87 (0.77–0.95) | 0.81 (0.69–0.90) | 0.83 (0.72–0.92) |
| | >100 | 0.64 | 0.75 | 0.64 | 0.97 | 0.84 | 0.97 | **0.85 (0.74–0.94)** | **0.84 (0.72–0.93)** | **0.82 (0.71–0.92)** |
| | >400 | 0.93 | 0.87 | 0.93 | 0.93 | 0.91 | 0.91 | 0.97 (0.91–1.00) | 0.94 (0.85–0.99) | 0.93 (0.84–0.98) |
| | >1000 | 1.00 | 1.00 | 1.00 | 0.91 | 0.94 | 0.96 | 0.97 (0.92–1.00) | 0.99 (0.95–1.00) | 0.99 (0.96–1.00) |

## 4. Discussion and Conclusion

This simulation-based feasibility study evaluated opportunistic CAC assessment using dual-layer DE chest radiography and a multi-task learning framework. Projection geometry strongly affected CAC conspicuity: the lateral view reduced overlap with dense bony anatomy and consistently produced higher Dice and AUC values than AP and 45° views. The AUC differences across projection angles did not reach statistical significance in the 59-patient test cohort after multiple-comparison correction, so these findings are interpreted as consistent performance trends rather than definitive superiority.

DE material decomposition provided the clearest statistically supported classification benefit. DE bone images significantly outperformed SE radiographs at CAC >100, >400, and >1000 after Holm-Bonferroni correction, whereas DE bone and HP bone did not differ significantly. Thus, material separation improved CAC discrimination, while additional high-pass filtering offered limited incremental benefit and became increasingly sensitive to noise at lower exposure.

Dose-dependent experiments showed a clearer decline in pixel-level localization than in case-level classification. Under matched-dose training and testing, no significant AUC differences were observed across 70, 50, and 24 mAs, whereas Dice decreased with exposure. The modest absolute Dice values should also be interpreted in the context of small-object segmentation: a 3-mm calcification projects to only about 6–7 pixels in diameter at the simulated geometry, so a one-pixel displacement can reduce Dice by approximately 18–21%. Accordingly, the segmentation branch is interpreted primarily as spatial localization and task-specific supervision, while score-based classification is the more clinically relevant screening endpoint.

Several limitations should be emphasized. First, the study uses a controlled CT-derived simulation rather than physical dual-layer radiographs. Although the detector geometry, scintillator configuration, MTF, and NNPS were experimentally constrained by the prototype system, scatter was not explicitly modeled and the results represent an idealized scatter-corrected condition, although many scatter correction methods exist that can yield accurate material decomposition[13,22]. Second, the ECG-gated CT-derived CAC reference and simulated radiographs share the same underlying anatomy and projection geometry, creating intentionally exact spatial correspondence. This is not a classical inverse-reconstruction setting, but it omits residual cardiac or respiratory motion, finite-exposure motion blur, registration uncertainty, and other model mismatch that may reduce localization performance in clinical imaging. Single-exposure dual-layer acquisition avoids inter-energy temporal misregistration, but does not eliminate all motion effects. Third, the held-out test cohort was limited to 59 patients, including only five positive cases at CAC >1000; bootstrap confidence intervals quantify uncertainty within this cohort but do not establish broader generalizability. Larger physical and clinical validation cohorts are therefore needed.

The score-regression head should also be interpreted as a learned CT-referenced estimator rather than an explicit implementation of the Agatston formula. Because the two softmax channels are complementary, global pooling effectively summarizes the continuous CAC probability burden, and the MLP learns a nonlinear calibration to the CT-derived score. Image features such as calcium conspicuity and morphology influence this probability map upstream through the nnU-Net, but lesion density is not entered as a separate explicit term.

In summary, lateral-view dual-layer DE bone imaging provides a promising configuration for opportunistic CAC assessment in this controlled simulation study. DE material decomposition showed a statistically supported advantage over SE radiography, while matched-dose score classification remained relatively stable across exposure levels. These findings establish feasibility and motivate larger physical and clinical validation before translation to routine chest-radiography workflows.